\documentclass[12pt,eqsecnum,preprint]{aastex}
\begin{document}
\newcommand{\el}{{\rm e}}
\newcommand{\xel}{x_{\rm e}}
\newcommand{\nel}{n_{\rm e}}
\newcommand{\h}{{\rm H}}
\newcommand{\mh}{m_{\rm H}}
\newcommand{\xh}{x({\rm H})}
\newcommand{\nh}{n_{\rm H}}
\newcommand{\Nh}{N_{\rm H}}
\newcommand{\hp}{{\rm H}^+}
\newcommand{\xhp}{x({\rm H}^+)}
\newcommand{\nhp}{n({\rm H}^+)}
\newcommand{\he}{{\rm He}}
\newcommand{\xhe}{x({\rm He})}
\newcommand{\nhe}{n({\rm He})}
\newcommand{\hep}{{\rm He}^+}
\newcommand{\xhep}{x({\rm He}^+)}
\newcommand{\nhep}{n({\rm He}^+)}
\newcommand{\neon}{{\rm Ne}}
\newcommand{\xne}{x({\rm Ne})}
\newcommand{\nne}{n({\rm Ne})}
\newcommand{\nep}{{\rm Ne}^+}
\newcommand{\xnep}{x({\rm Ne}^+)}
\newcommand{\nnep}{n({\rm Ne}^+)}
\newcommand{\hm}{{\rm H}_2}
\newcommand{\xhm}{x({\rm H}_2)}
\newcommand{\nhm}{n({\rm H}_2)}
\newcommand{\hmp}{{\rm H}_{2}^{+}}
\newcommand{\xmp}{x({\rm H}_{2}^{+})}
\newcommand{\xhmp}{x({\rm H}_{2}^{+})}
\newcommand{\nhmp}{n({\rm H}_{2}^{+})}
\newcommand{\hthreep}{{\rm H}_{3}^{+}}
\newcommand{\xhthreep}{x({\rm H}_{3}^{+})}
\newcommand{\nhthreep}{n({\rm H}_{3}^{+})}
\newcommand{\hmin}{{\rm H}^-}
\newcommand{\xhmin}{x({\rm H}^-)}
\newcommand{\nhmin}{n({\rm H}^-)}
\newcommand{\xc}{x({\rm C})}
\newcommand{\nc}{n({\rm C})}
\newcommand{\cp}{{\rm C}^+}
\newcommand{\xcp}{x({\rm C}^+)}
\newcommand{\ncp}{n({\rm C}^+)}
\newcommand{\kc}{k_{\rm C}}
\newcommand{\xo}{x({\rm O})}
\newcommand{\no}{n({\rm O})}
\newcommand{\op}{{\rm O}^+}
\newcommand{\xop}{x({\rm O}^+)}
\newcommand{\nop}{n({\rm O}^+)}
\newcommand{\xn}{x({\rm N})}
\newcommand{\nn}{n({\rm N})}
\newcommand{\np}{{\rm N}^+}
\newcommand{\nnp}{n({\rm N}^+)}
\newcommand{\xnp}{x({\rm N}^+)}
\newcommand{\ntwo}{{\rm N}_{2}}
\newcommand{\xntwo}{x({\rm N}_{2})}
\newcommand{\nntwo}{n({\rm N}_{2})}
\newcommand{\ntwop}{{\rm N}_{2}^+}
\newcommand{\xntwop}{{x(\rm N}_{2}^+)}
\newcommand{\nntwop}{{n(\rm N}_{2}^+)}
\newcommand{\ntwohp}{{\rm N}_{2}{\rm H}^+}
\newcommand{\xntwohp}{{x(\rm N}_{2}{\rm H}^+)}
\newcommand{\nntwohp}{{x(\rm N}_{2}{\rm H}^+)}
\newcommand{\xna}{x({\rm Na})}
\newcommand{\nna}{n({\rm Na})}
\newcommand{\nap}{{\rm Na}^+}
\newcommand{\xnap}{x({\rm Na}^+)}
\newcommand{\nnap}{n({\rm Na}^+)}
\newcommand{\xa}{x({\rm A})}
\newcommand{\ap}{{\rm A}^+}
\newcommand{\xap}{x({\rm A}^+)}
\newcommand{\ka}{k_{\rm A}}
\newcommand{\xm}{x({\rm m})}
\newcommand{\nm}{n({\rm m})}
\newcommand{\km}{k_{\rm m}}
\newcommand{\oh}{{\rm OH}}
\newcommand{\xoh}{x({\rm OH})}
\newcommand{\noh}{n({\rm OH})}
\newcommand{\ohp}{{\rm OH}^+}
\newcommand{\xohp}{x({\rm OH}^+)}
\newcommand{\nohp}{n({\rm OH}^+)}
\newcommand{\htwoo}{{\rm H}_{2}{\rm O}}
\newcommand{\xhtwoo}{x({\rm H}_{2}{\rm O})}
\newcommand{\nhtwoo}{n({\rm H}_{2}{\rm O})}
\newcommand{\htwoop}{{\rm H}_{2}{\rm O}^{+}}
\newcommand{\xhtwoop}{x({\rm H}_{2}{\rm O}^{+}}
\newcommand{\nhtwoop}{n({\rm H}_2{\rm O}^+)}
\newcommand{\hthreeop}{{\rm H}_{3}{\rm O}^{+}}
\newcommand{\xhthreeop}{x({\rm H}_{3}{\rm O}^{+})}
\newcommand{\nhthreeop}{n({\rm H}_{3}{\rm O}^{+})}
\newcommand{\otwo}{{\rm O}_2}
\newcommand{\xotwo}{x({\rm O}_2)}
\newcommand{\notwo}{n({\rm O}_2)}
\newcommand{\otwop}{{\rm O}_{2}^{+}}
\newcommand{\xotwop}{x({\rm O}_{2}^{+})}
\newcommand{\notwop}{n({\rm O}_{2}^{+})}
\newcommand{\otwoh}{{\rm O}_2{\rm H}}
\newcommand{\xotwoh}{x({\rm O}_2){\rm H}}
\newcommand{\notwoh}{n({\rm O}_2){\rm H}}
\newcommand{\otwohp}{{\rm O}_2{\rm H}^+}
\newcommand{\xotwohp}{x({\rm O}_2){\rm H}^+}
\newcommand{\notwohp}{n({\rm O}_2){\rm H}^+}
\newcommand{\co}{{\rm CO}}
\newcommand{\xco}{x({\rm CO})}
\newcommand{\nco}{n({\rm CO})}
\newcommand{\Nco}{N({\rm CO})}
\newcommand{\xcop}{x({\rm CO}^+)}
\newcommand{\ncop}{n{(\rm CO}^+)}
\newcommand{\hcop}{{\rm HCO}^+}
\newcommand{\xhcop}{x({\rm HCO}^+)}
\newcommand{\nhcop}{n{(\rm HCO}^+)}
\newcommand{\Mst}{M_{\ast}}
\newcommand{\Tst}{T_{\ast}}
\newcommand{\Rst}{R_{\ast}}
\newcommand{\LX}{L_{\rm bol}}
\newcommand{\Lbol}{L_{\rm X}}
\newcommand{\TX}{T_{\rm X}}
\newcommand{\tauX}{\tau_{\rm X}}
\newcommand{\Rx}{R_{\rm x}}
\newcommand{\Ox}{\Omega_{\rm x}}
\newcommand{\Mdotw}{\dot{M}_{\rm W}}
\newcommand{\nw}{n_{\rm W}}
\newcommand{\rhow}{\rho_{\rm W}}
\newcommand{\Tw}{T_{\rm W}}
\newcommand{\pw}{p_{\rm W}}
\newcommand{\vw}{v_{\rm W}}
\def\ppd{protoplanetary disk\,}
\def\ppds{protoplanetary disks\,}
\newcommand{\Nperp}{N_{\perp}}
\newcommand{\alphah}{\alpha_{\rm h}}
\newcommand{\Td}{T_{\rm d}}
\newcommand{\Tg}{T_{\rm g}}
\newcommand{\nd}{n_{\rm d}}
\newcommand{\rhod}{\rho_{\rm d}}
\newcommand{\rhog}{\rho_{\rm g}}
\newcommand{\Tda}{T_{\rm d}(a)}
\newcommand{\nda}{n_{\rm d}(a)}
\newcommand{\Mdot}{\dot{M}}
\newcommand{\zetacr}{\zeta_{\rm CR}}
\newcommand{\zetax}{\zeta_{\rm X}}
\newcommand{\Msun}{M_{\odot}}
\newcommand{\Rsun}{R_{\odot}}\newcommand{\pcc}{{\rm cm}^{-3}}
\newcommand{\percc}{\rm \,cm^{-3}}
\newcommand{\psqcm}{{\rm cm}^{-2}}
\newcommand{\persqcm}{\rm \,cm^{-2}}
\newcommand{\gpersqcm}{\rm \,g\,cm^{-2}}
\newcommand{\gpercc}{\rm \,g\,cm^{-3}}
\newcommand{\ps}{{\rm s}^{-1}}
\newcommand{\ccps}{{\rm cm}^{3} {\rm s}^{-1}}
\newcommand{\kmps}{{\rm km}\,{\rm s}^{-1}}
\newcommand{\erg}{{\rm erg}}
\newcommand{\kb}{k_{\rm B}}
\def\micron{\hbox{$\mu$m}}
\newcommand{\bcen}{\begin{center}}
\newcommand{\ecen}{\end{center}}
\newcommand{\be}{\begin{equation}}
\newcommand{\ee}{\end{equation}}
\newcommand{\bdis}{\begin{displaymath}}
\newcommand{\edis}{\end{displaymath}}
\def\eg{{e.g.\ }}
\def\etc{{etc.\ }}
\def\etal{\mbox{\it et al.\,}}
\def\ie{{i.e.\ }}
\def\noi{{\noindent}}
\def\ra{{\rightarrow}}

\slugcomment{Accepted by ApJ October 8, 2006}
\shorttitle{Neon Fine Structure Lines}
\shortauthors{Glassgold, Najita, and Igea}

\title{Neon Fine-Structure Line Emission By 
X-ray Irradiated Protoplanetary Disks}

\author{Alfred E. Glassgold}
\affil{Astronomy Department, University of California, Berkeley, CA 94720}
\email{aglassgold@astron.berkeley.edu}

\author{Joan R. Najita}
\affil{NOAO, 950 N.~Cherry Avenue, Tucson, AZ 85719}
\email{najita@noao.edu}

\author{Javier Igea}
\affil{Vatican Observatory, V-00120 Citt\'a del Vaticano}
\email{jigea@telefonica.net}


\begin{abstract}
Using a thermal-chemical model for the generic T-Tauri disk of
D'Alessio \etal (1999), we estimate the strength of the fine-structure
emission lines of Ne\,II and Ne\,III at 12.81 and 15.55 $\micron$ that
arise from the warm atmosphere of the disk exposed to hard stellar
X-rays. The Ne ions are produced by the absorption of keV X-rays from
the K shell of neutral Ne, followed by the Auger ejection of several
additional electrons. The recombination cascade of the Ne ions is slow
because of weak charge transfer with atomic hydrogen in the case of
Ne$^{++}$ and by essentially no charge transfer for Ne$^+$. For a
distance of 140\,pc, the 12.81\,$\micron$ line of Ne\,II has a flux
$\sim 10^{-14} \erg\, \psqcm \ps$, which should be observable with the
{\it Spitzer} Infrared Spectrometer and suitable ground based
instrumentation. The detection of these fine-structure lines would
clearly demonstrate the effects of X-rays on the physical and chemical
properties of the disks of young stellar objects and provide a
diagnostic of the warm gas in protoplanetary disk atmospheres. They
would complement the observed $\hm$ and CO emission by probing
vertical heights above the molecular transition layer and larger
radial distances that include the location of terrestrial and giant
planets.

\end{abstract}


\section{Introduction}
 
Young stellar objects (YSOs) are observed to be strong emitters of
X-rays (e.g., Feigelson \& Montmerle 1999). The X-rays have the
potential to affect the physical properties of nearby circumstellar
material (e.g., Glassgold, Feigelson \& Montmerle 2000) and thus the
course of star formation (Silk \& Norman 1980; Pudritz \& Silk
1987). Tsujimoto \etal (2005) observed X-ray fluorescence from cold Fe
ions in seven sources in the Orion Nebula Cluster, which they have
interpreted to be the result of absorption of hard stellar X-rays by
cool circumstellar disk material. This confirmation of the interaction
of stellar X-rays with the disks of YSOs raises the question of
whether there are specific diagnostics for the effect of X-rays on the
physical and chemical properties of these disks, which are believed to
play a crucial role in the formation of stars and planets.  Although
the gas is the main reservoir of mass for much of the lifetime of a
YSO disk, much less is known about the gas than the easier to measure
dust. Thus we address the issue of X-ray effects in the broader
context of the search for gaseous diagnostics of disk gas.

This type of question has a long history in studies of the
interstellar medium of our own and external galaxies, where definitive
demonstrations of the effects of X-rays have been elusive because
other external heating and ionizing radiations, such as UV radiation
and cosmic rays, can produce similar effects. Some of the diagnostic
possibilities discussed in the literature are multiply charged ions
(Dalgarno 1976; Langer 1978); specific molecules, especially ions and
radicals (Krolik \& Kallmann 1983; Neufeld, Maloney \& Conger 1994;
Lepp \& Dalgarno 1996; Sternberg, Yan \& Dalgarno 1997; Yan \&
Dalgarno 1997); near-infrared H$_2$ ro-vibrational transitions (Lepp
\& McCray 1983; Draine \& Woods 1989; Gredel \& Dalgarno 1995; Tin\'e
\etal 1997); and fine-structure lines (Maloney, Hollenbach \& Tielens
1996).

Many of these possibilities are now being re-investigated in the
context of chemical modeling of the disks and envelopes of YSOs
irradiated by X-rays and UV radiation (e.g., Aikawa \& Herbst 1999;
Markwick \etal 2002; Gorti \& Hollenbach 2004; Millar \& Nomura 2005;
St\"auber \etal 2005). An important aspect of the X-ray irradiation of
protoplanetary disks is that the upper atmosphere close to the YSO is
heated to relatively high temperatures $\sim 5,000$\,K (Glassgold,
Najita \& Igea 2004, henceforth GNI04; Alexander, Clarke \& Pringle
2004). The temperature then decreases going down towards the disk
midplane. The accompanying thermal-chemical structure consists of a
hot partially-ionized atomic layer on top of a cool molecular layer
with a much smaller ionization fraction. In between, there is a warm
partially-molecular layer with gas temperatures $\sim
500-2,000$\,K. This layered structure of X-ray irradiated disks can be
traced back to earlier studies of various astrophysical environments,
e.g., Halpern \& Grindlay (1983), Krolik \& Kallman (1983) and Lepp \&
McCray (1983). The thermal properties of the disk atmosphere above the
cool molecular layer play a critical role in determining the
observational signatures of disks.

The potential for ambiguity in the identification of X-ray diagnostics
also arises in protoplanetary disks because of the presence of stellar
UV radiation which can heat the gas in the surface layers via the
photoelectric effect on small grains and polycyclic aromatic
hydrocarbons compounds or PAHs (Kamp \& Dullemond 2004; Nomura \&
Millar 2005). In the far outer regions of disks, interstellar UV
radiation and galactic cosmic rays may also play a role in heating and
ionizing the gas. We suggest that the mid-infrared fine-structure
lines of Ne\,II at 12.81 $\mu$m and of Ne\,III at 15.6 $\mu$m may
provide unique signatures of X-ray ionization of the disks of low-mass
YSOs.

Neon fine-structure lines have proved invaluable for studying ionized
regions in our own as well as external galaxies (e.g., see the recent
observations by Jaffe \etal 2003 of ultra compact H\,II regions in
Monoceros R2, by Sturm \etal 2001 of AGN, by Devost \etal 2004 of the
starburst galaxy NGC\,253 and by Weedman \etal 2005 of Seyfert
galaxies). The Ne ions in H\,II regions and galaxies are generated by
the Lyman continuum photons from massive stars and by
X-rays. Generating these ions with UV photons is much more difficult
in the environment of low-mass young stellar objects. Cosmic rays are
also unlikely to be effective for this purpose because they are
excluded by the strong winds generated by YSOs (Glassgold, Najita \&
Igea 1997).

Low-mass YSOs are known to be strong emitters of moderately hard keV
X-rays (e.g., Feigelson \& Montmerle 1999), and we suggest that the Ne
ions are mainly produced by X-rays. The basic idea is that the high
ionization potentials of Ne and Ne$^+$ (21.56 and 41.0\,eV,
respectively) preclude the production of these ions by the FUV
radiation ($\lambda > 911.6$\,\AA \,) emitted by T Tauri stars. We
will show that the hard X-rays emitted by YSOs are effective in
generating Ne$^+$ and Ne$^{2+}$ over a range of radii in the
atmospheres of the disks of these stars. The emission arises in the
same region of the disk from which photo-evaporation is believed
to originate. Thus the Ne ion fine-structure line emission can serve as
a diagnostic of the source of an evaporative flow as well as the
signature of X-ray irradiation.

T Tauri stars are also likely to emit significant amounts of EUV
radiation ($\lambda \lesssim 911.6$\,\AA \,) (e.g., Alexander, Clarke
and Pringle 2005), although it is much less well characterized than
their X-ray emission. EUV radiation has been invoked to
photo-evaporate disks (reviewed by Hollenbach \etal 2000, Dullemond
\etal 2006). However, EUV photons are rapidly absorbed by atomic
hydrogen in the partially-ionized wind and disk of YSOs (Alexander,
Clarke and Pringle 2003) so that, except for a fully-ionized
region near the star (Hollenbach \etal 1994), the EUV flux
that reaches substantial radii and depths in disks to generate Ne ions
is probably very small. The small ionized region generated by stellar EUV radiation may be the source of some Ne
fine-structure line emission (U. Gorti and D. Hollenbach, private
communication 2006).

The plan of this article is as follows. First, we lay the foundation
for calculating the fluxes of the X-ray generated Ne fine-structure
emission by developing the ionization theory for the Ne ions.  We take
into account production by X-rays and destruction by charge exchange
as well as radiative recombination in a form appropriate to the
thermal-chemical structure of X-ray irradiated protoplanetary disk
atmospheres. In \S 3, we calculate the fluxes of the fine-structure
lines for the disk atmosphere of a typical T Tauri disk, where the
electron densities are generally sub-critical. In \S 4 we discuss the
observability of these lines, their potential astrophysical
significance, and the uncertainties in the theory.

\section{Ionization Theory} 

We start by discussing the production and destruction of Ne ions in
the inner region of a \ppd \,atmosphere, within tens of AU. We
consider only X-rays for ionizing the neutral and low-ionization forms
of the Ne atom, since FUV photons have insufficient energy to ionize
them, EUV is absorbed over short distances in mainly neutral matter,
and cosmic rays are largely excluded from the inner regions, as
mentioned in \S 1.  Collisional ionization by thermal electrons is
unimportant for the relevant temperatures, $T \lesssim 10,000$\,K.

The result of an X-ray absorption by a low-ionized Ne ion depends on
whether the X-ray energy $E$ is less than or greater than the K-shell
threshold, $E_K = 0.870, 0.903,\, \ldots \,$\,keV for Ne\,I, Ne\,II \,
$\ldots$ \,. According to simple one-electron theory (e.g., Kaastra \&
Mewe 1993), an L-shell electron is ejected when $E < E_K$, increasing
the ion charge by one. For $E > E_K$, the cross section for L-shell
photoionization is small compared with the production of a K-shell
vacancy, and the latter leads mainly to the ejection of another
electron by the Auger process. According to one-electron theory
(ignoring correlations), the branching ratio for single ionization is
only $2\%$ (Kaastra \& Mewe 1993). The actual situation is more
complicated, especially with regard to the production of more than two
electrons, as was shown in laboratory experiments on neutral Ne by
Carlson \& Krause and collaborators (Krause \etal 1964; Carlson \&
Krause 1965a,b). First of all, there is a small but significant
probability, $\sim 10\%$, that L-shell photoionization ejects {\it
two} electrons. Second, the probability that K-shell vacancies produce
just one electron is somewhat larger ($\sim 6\%$) than predicted by
one-electron theory ignoring electron correlations. Third, and most
important, $\sim 30\%$ of K-shell vacancies produced in the
photoionization of neutral Ne lead with significant probability to
higher ions such as Ne$^{3+}$ and Ne$^{4+}$. However, in
partially-ionized regions characteristic of the atmospheres of \ppds,
where atomic hydrogen is the dominant species in contrast to fully
ionized H\,II regions, Ne$^{3+}$ is rapidly destroyed by charge
exchange with atomic hydrogen,
\be
{\rm Ne}^{3+} + {\rm H} \rightarrow  {\rm Ne}^{2+} + \hp.       
\ee                             
The cross section for this reaction has been measured to be large and
rising down to 0.1\,eV by Rejoub \etal (2004); it corresponds to rate
coefficients for charge transfer $\sim 10^{-9}\, \ccps$. In essence,
ions beyond Ne$^{2+}$ that arise from K-shell vacancies in Ne atoms
and low-ionization ions are rapidly converted back to Ne$^{2+}$ by
one or more charge transfers with atomic hydrogen.

Charge exchange with atomic hydrogen is also important in the
ionization balance of Ne, Ne$^+$ and Ne$^{2+}$. Charge transfer of
Ne$^{2+}$ to H was studied by Dalgarno and collaborators 25 years ago
(Butler, Bender \& Dalgarno 1979; Dalgarno, Butler \& Heil 1980;
Butler, Heil \& Dalgarno 1980; Butler \& Dalgarno 1980). Their
conclusion was that the rate coefficient for this process is
exceedingly small, but that {\it radiative} charge exchange takes
place with rate coefficient $k_2 \approx 10^{-14} \ccps$. Although
this is much smaller than the (essentially Langevin) rate coefficient
for fast charge transfer ($\sim 10^{-9}\, \ccps$), it cannot be
ignored in partially-ionized regions because its effect is the same
order of magnitude as radiative recombination, $\propto \xel
\alpha_2$, where $\alpha_2$ is the recombination rate coefficient for
Ne$^{2+}$ and $\xel$ is the fractional ionization. There are no
reports on charge transfer from Ne$^+$ to H, but the potential energy
curves calculated for this system by Pendergast, Heck and Hayes (1994)
indicate that radiative charge exchange is highly forbidden
(A. Dalgarno, private communication). We conclude that Ne$^+$ is
primarily destroyed by radiative recombination and, to a lesser
extent, by X-ray photoionization into higher ions.

These considerations on Ne$^+$ and Ne$^{2+}$ charge transfer to H
provide the basis for a simple set of steady-state ionization balance
equations. The steady state assumption is justified by the high electron
density in the region relevant for the emission of the fine-structure 
lines, typically $\sim 10^5 \pcc$. 
At these high densities, 
the time scale for the slowest process, radiative recombination, is $\sim 1$\, yr, the same order or less than the orbital time and, more important, significantly less than the accretion time scale. 
We denote the fraction of Ne in charge state $n$ by $x_n$
($n=0,1,2$). We express the difference between X-ray ionization below
and above the Ne\,I K-edge by writing the total X-ray ionization rate
as the sum of a soft and a hard component,
\be
\zeta(n)  = \zeta_s(n) + \zeta_h(n) \hspace{0.75in}(n=0,1).
\ee
Following the above discussion, we introduce branching ratios $B_1$
and $B_2$ for soft X-ray ionization of Ne to produce Ne$^+$ ($B_1
\approx 0.9$) and for hard X-ray ionization of Ne to produce Ne$^{2+}$
($B_2 \approx 0.94$, when fast charge exchange of higher ions back to
Ne$^{2+}$ is included). The steady balance equations for Ne$^+$ and
Ne$^{2+}$ are then:
\be
\left [B_1 \frac{\zeta_s(0)}{\nh} + 
(1-B_2) \frac{\zeta_h(0)}{\nh} \right ] x_0 
+ (\alpha_2 \xel + k_2)x_2 = 
\left [\frac{\zeta(1)}{\nh} + (\alpha_1 \xel + k_1) \right ] \, x_1,
\ee    
\be
\left [(1-B_1)\frac{\zeta_s(0)}{\nh} + 
B_2\frac{\zeta_h(0)}{\nh}\right ] \, x_0 
+ \frac{\zeta(1)}{\nh}\, x_1 = (\alpha_2 \xel + k_2)\, x_2,
\ee
where $\nh$ is the number density of hydrogen nuclei. These equations
can easily be solved in terms of the ion abundance ratios,
\be
\frac{x_1}{x_0} = \frac{\zeta(0)/\nh}{\alpha_1 \xel + k_1},
\ee
\be
\frac{x_2}{x_0}= 
B_2\frac{\zeta_h(0)/\nh}{\alpha_2 \xel + k_2} + 
(1-B_1)\frac{\zeta_s(0)/\nh}{\alpha_2 \xel + k_2} + 
\frac{\zeta(1)/\nh}{\alpha_2 \xel + k_2} \, \frac{x_1}{x_0},
\ee
with the proviso that $x_0 + x_1 + x_2 = x_{\rm Ne}$, where $x_{\rm
Ne}$ is the disk abundance of gaseous Ne.  When we recall from \S 2 
that $k_1$ is negligible and ignore the small quantities
$(1-B_n)$ and the small difference between $\zeta(0)$ and
$\zeta(1)$, we obtain the following relations:
\be
\frac{x_1}{x_0} = \frac{\zeta({\rm Ne})  /\nh}{\alpha_1 \xel},
\ee
\be
\frac{x_2}{x_0}= 
\frac{\zeta_h({\rm Ne})  /\nh}{\alpha_2 \xel + k_2} + 
 \frac{\zeta({\rm Ne}) /\nh}{\alpha_2 \xel + k_2}\, \frac{x_1}{x_0},
\ee
where $\zeta({\rm Ne}) = \zeta_s({\rm Ne}) + \zeta_h({\rm Ne})$ is the
ionization rate for neutral Ne per Ne nucleus. The direct X-ray
ionization rates for Ne can be calculated from the absorption cross
section (Morrison \& McCammon 1983; Wilms \etal 2000). Here we adopt
an approximate treatment in which the Ne rates are scaled from
$\zeta$, the ionization rate per H nucleus, according to the ratio of
the Ne cross section to the mean cross section that defines $\zeta$.
Following GNI04, we assume that the X-ray luminosity and spectral
temperature are $\LX = 2\times 10^{30}$\,erg\,$\ps$ and $k\TX =
1$\,keV, so that the X-ray spectrum extends from several hundred eV to
a few KeV. Good representative figures before attenuation are
$\zeta({\rm Ne}) \approx 40 \zeta$ and $\zeta_h({\rm Ne}) / \zeta({\rm
Ne}) \approx 7/8$. These ratios are actually a function of position
because the X-ray spectrum hardens due to the preferential absorption
of soft X-rays. Thus the ratio $\zeta_h({\rm Ne}) / \zeta({\rm Ne})$
tends to unity at large column density.  The ratio $\zeta({\rm Ne}) /
\zeta $ is much greater than one because the X-rays emitted by young
stellar objects have characteristic energies of the same order or
larger than the Ne K-edge near 0.9\,keV, where the Ne absorption cross
section is maximum. In particular, the photoionization cross section
of Ne for keV X-rays is much larger than the cross section summed over
a cosmic abundance mix (Morrison and McCammon 1983; Wilms \etal 2000)
that determines $\zeta$, the ionization rate per hydrogen
nucleus. If the stellar X-ray spectrum contained few keV photons, the
ratios $\zeta({\rm Ne}) / \zeta$ and $\zeta_h({\rm Ne}) / \zeta({\rm
Ne})$ would be very much smaller than the representative values quoted
above. The secondary electrons generated by the primary
photoionization can ionize hydrogen as well as a heavy element like Ne. But
the electronic ionization cross section for Ne is generally no more
than twice those for hydrogen and helium. Following the discussion of
Maloney, Hollenbach \& Tielens (1996), we estimate that the
contribution of the secondary electrons to the X-ray ionization rate
of Ne is $\lesssim 2 \zeta$, and thus small compared to the direct X-ray
ionization.

We will find in \S 3 that, for the regions of the disk responsible for
the emission of the fine-structure lines, the fractions $x_1/x_0$ and
$x_2/x_0$ are usually $\lesssim 0.1$ and thus
\be
\label{ion_ratios}
\frac{x_1}{x_0} = \frac{\zeta({\rm Ne}) /\nh}{\alpha_1 \xel} 
\equiv a_1,
\hspace{0.75in}
\frac{x_2}{x_0}\approx \frac{\zeta_h ({\rm Ne}) /\nh}{k_2 +\alpha_2 \, \xel} 
\equiv a_2.
\ee
Because $\alpha_2 > \alpha_1$ and $k_2$ is finite though small, these
equations tell us that there is substantially more Ne$^+$ than
Ne$^{2+}$.

Over most of the region of interest, the electron fraction is
approximately given in terms of the total ionization rate by the
simple formula,
\be
\label{simple_ionization}
\xel \approx \left ( \frac{\zeta}{\nh \alpha({\rm H}^{+})} \right )^{1/2},
\ee  
where $\alpha({\rm H}^{+})$ is the total recombination rate of
$\hp$. Therefore, according to equations~\ref{ion_ratios} and
\ref{simple_ionization}, the fractions of Ne in the first two ion
states depend on different powers of the ionization parameter $\zeta /
\nh$ and, equivalently, the electron fraction,
\be
\label{chargeratios}
\frac{x_1}{x_0} \sim (\zeta /\nh)^{1/2} \sim \xel,
\hspace{0.75in}
\frac{x_2}{x_0} \sim \zeta /\nh \sim \xel^2 .
\ee
The actual ionization fractions used in the calculations of \S3 are,
\be
\label{actualpops}
x_0/x_{\rm Ne} = \frac{1}{d}, 	\hspace{0.25in}
x_1/x_{\rm Ne} = \frac{a_1}{d},	\hspace{0.25in}	
x_2/x_{\rm Ne} = \frac{(a_1 + b)a_2}{d},
\ee
where $d=1 + a_1 + (a_1+b)a_2$ with $b \equiv \zeta_h({\rm Ne})/
\zeta({\rm Ne}) \approxeq 0.875$. 

In addition to their utility in the numerical calculations, these
analytic formulae provide some insight into the physics of Neon
ionization in X-ray irradiated protoplanetary disk atmospheres.  In
the first member of Eq.~\ref{chargeratios}, the approximate
proportionality of $x_1/x_0$ to the electron fraction is similar in
form to the well known ratio of O$^+$ to O when O$^+$ and H$^+$ are
strongly coupled by fast, near-resonant, charge exchange (e.g.,
Osterbrock 1987). In the present case, Ne$^+$ and H$^+$ are not
connected directly by charge exchange, but they are linked by both
being produced by X-rays. A semi-quantitative expression of this
relation, based on Eq.~\ref{chargeratios}, is,
\be
\label{neplusratio}
\frac{x({\rm Ne}^+)}{x({\rm Ne})} 
\sim \frac{\alpha({\rm H}^{+})}{\alpha_1}\,
\frac{\zeta({\rm Ne})}{\zeta}\, \xel.
\ee   
Because the ratio $\zeta({\rm Ne}) / \zeta \gg1$ and $\alpha_1$ is not
very much smaller than $\alpha({\rm H}^{+})$, $x({\rm Ne}^+) /x({\rm
Ne})$ is much larger than $\xel$.  Although Eq.~\ref{neplusratio}
expresses much of the essential physics of the X-ray ionization of Ne
for protoplanetary disks, it is inadequate for quantitative estimates
because it breaks down for very small and very large $\xel$.


In the next section, we use the detailed Ne ionization theory 
presented above to estimate the Ne ion line fluxes
for a T-Tauri disk.  We adopt, for this purpose, the thermal-chemical 
disk structure of an existing model that includes 
stellar X-ray heating and ionization (GNI04) but which did
not include the Ne ionization theory discussed above.
The GNI04 model allows for the inclusion
of additional mechanical heating of the disk atmosphere, e.g., from accretion or wind-disk interaction, using the following formula for the volumetric heating rate
\be
\label{accheat}
\Gamma_{\rm mech} = \frac{9}{4} \alpha_h \rho c^2 \Omega, 
\ee 
where $\alpha_h $ is a phenomenological parameter. Figure 1 is an
example of the results of GNI04 at a radius of $R=5$\,AU; additional
results for a range of radii are presented below in Figures 2 and
3. Figure 1 compares the vertical variation of the gas and dust
temperatures. The abscissa represents the height above the midplane
$z$ in terms of the vertical column density $\Nh$ of hydrogen from
$\infty$ down to $z$. The calculations are based on the D'Alessio
\etal (1999) fiducial dust model for a T Tauri star with mass $\Mst=
0.5 \Msun$, radius $\Rst = 2 \Rsun$, and effective temperature $\Tst =
4000$\,K. The accretion rate is $\dot{M} = 10^{-8} \Msun$ yr$^{-1}$,
and the disk thickness at 1\,AU is $\sim 100$ \,g $\psqcm$ and varies
elsewhere roughly as $1/R$.  Heating of small grains by the stellar
optical and near infrared radiation produces an inversion in the dust
temperature $T_d$ over the midplane value, as shown by the dashed
curve. X-ray heating generates a much larger inversion in the {\it
gas} temperature $T_g$ (solid line labeled $\alpha_h = 0.01$) in the
region above the warm dust. When mechanical heating is included (the
solid curve labeled $\alpha_h = 1.0$), the hot gas extends even
further down into the disk. The gas and dust temperatures approach one
another near the midplane (large $\Nh$), where the volume densities
$\nh$ are big enough to ensure good thermal coupling between the gas
and the dust. In between the hot gas layer at the top of the
atmosphere and the relatively cool midplane region is a warm
transition zone ($\Nh \gtrsim 10^{21}\psqcm$) where $T_g \approx
500-2000$\,K. Gas temperature inversions can also be obtained with UV
heating of small grains and PAHs (Kamp \& Dullemond 2005, Nomura \&
Millar 2005). Major chemical transitions occur between $10^{21}$ and
$10^{22}\, {\rm cm}^{-2}$. Significant levels of $\hm$ and full
formation of CO occur near $\Nh \sim 1-2 \times 10^{21} {\rm
cm}^{-2}$, and complete formation of $\hm$ is achieved at $\Nh = 8.5
\times 10^{21}\, {\rm cm}^{-2}$.

Figure 2 shows gas temperature profiles for radii ranging from
$R=1-40$\,AU for the case $\alpha_h = 0.01$. They exhibit a hot gas
layer on top of a cool one, with temperatures generally decreasing
with increasing radial distance. The temperature at the top of the
atmosphere decreases with radius due to inverse square dilution of the
stellar X-rays. The temperature near the midplane decreases with
distance because the viscous heating in the D'Alessio model
decreases. The results for the case where mechanical heating dominates
are similar, except that the transition from hot to cold occurs deeper
down in the atmosphere. Beyond $R=20$\,AU, there is a qualitative
change in the temperature profiles in that the maximum temperature
drops from several thousand to several hundred K. Figure 3 shows
clearly how X-ray irradiation can warm disk atmospheres out to large
radii of the order of tens of AU.  It is also noteworthy that the
near-midplane temperature at large radii can fall below the grain
freeze-out temperature for many volatile species. Although the
freeze-out process is omitted from our present chemical model, it does
not affect the calculation of the Ne fine-structure emission, which
arises from the upper atmosphere of the disk where the temperature is
high and freeze-out does not occur. Regions where freeze-out can occur
do not contribute to the emission because the upper levels of the
transitions, with excitation temperatures $\sim 1000$\,K, are
difficult to excite with electron collisions at low temperatures.

Figure 3 shows that the electron fraction decreases smoothly at high
altitudes. It then drops more rapidly to a level $\xel \sim 10^{-4} -
10^{-5}$, at a depth close to where $T_g$ begins to approach $T_d$ and
near where carbon becomes almost fully associated into CO. Figure 3
demonstrates that stellar X-rays can generate substantial levels of
ionization in disk atmospheres out to large radii. The variation in
gas temperature and ionization in Figures 2 and 3 are directly
relevant to the strength of the Ne line fluxes. Above the
thermal-chemical transition, especially in the range near $\Nh =
10^{19} - 10^{20}\psqcm$, the warm ionized conditions are conducive to
obtaining large Ne ion abundances and significant populations of the
upper levels of the Ne\.II 12.81 $\mu$m and Ne\,III 15.55 $\mu$m
\,transitions.

\section{Flux Calculation}

We estimate the emission from the Ne\,I and Ne\,II fine-structure
lines from a face-on disk in the optically-thin approximation.  The
lines originate in magnetic dipole transitions from the Ne\,II doublet
and the Ne\,III triplet, whose properties are given in Table 1.
\clearpage
\begin{table}
\begin{center} 
\caption{Ne Fine Structure Levels}
\begin{tabular}{ccccccc}
\hline 
\hline 
Ion & $\lambda$\,($\mu$m) &$J_u - J_l$&$T_{ul}$(K)& $A_{ul}$ (s$^{-1}$) & $T^{-1/2}
n_{cr}(ul)$ ($\pcc$)& $A_{ul}h\nu_{ul}$ (erg s$^{-1}$)\\ 
\hline 
Ne$^+$ & 12.81
& 1/2-3/2 & 1122.8 & $8.59 \times 10^{-3}$  & $5.53 \times 10^3$ &
$1.332\times 10^{-15}$ \\ Ne$^{2+}$ & 15.55 & 1-2 & 925.8 & $5.97
\times 10^{-3}$ & $3.94\times 10^3$ & $7.629\times 10^{-16}$\\
Ne$^{2+}$ & 36.02 & 0-1 & 399 & $1.15 \times 10^{-3}$ & $7.20\times
10^2$ & $6.337\times 10^{-17}$\\ 
\hline 
\end{tabular} 
\end{center} 
\end{table}
\clearpage
The critical electron densities for the transitions ($n_{cr} = A_{ul}
/ k_{ul}$, where $k_{ul}$ is the electronic collisional de-excitation
rate coefficient) are taken from Mendoza (1983) and are given in the
next to last column of Table 1. We assume that electronic collisions are more important than H collisions in exciting these fine-structure transitions. 

We ignore the $J=0-1$ 36.02 $\mu$m transition of Ne\,III, which is an
order of magnitude weaker than the $J=1-2$ 15.55 $\mu$m transition,
and the very weak $0-2$ magnetic quadrupole transition (not listed in
Table 1). The optically-thin approximation is valid because the
emission comes from a surface layer with limited vertical column
density $\Nh <10^{22}\, \psqcm$. Deeper regions ($\Nh > 10^{22}\,
\psqcm$) do not contribute because they are either insufficiently
ionized or too cool for the transitions to be collisionally excited.

The flux in a transition $u \ra l$ of an ion with charge $n$ (the ion
label $n=1,2$ now becomes a line label, $n=1$ for the 12.81 and $n=2$
for the 15.55 $\mu$m transition) from a disk annulus between $R$ and
$R+dR$ is,
\be 
\label{differential_flux} 
dF_{ul}(R;n) = \frac{1}{4 \pi
d^2}\, 2\pi R dR \int_0^\infty dN\, P_{u}(n)\, x_n \, (A_{ul}h
\nu_{ul})_n \, , 
\ee 
where $P_{u}(n)$ is the normalized population of the upper level,
$A_{ul}$ and $h \nu_{ul}$ are the usual $A$-value and energy of the
transition, and $dN = d\Nh$ is the differential of the vertical
hydrogen column density. At high densities, the populations are given
by the Boltzmann distribution, whereas at low densities they reduce to
the probability for a collisional excitation to occur within the
radiative lifetime of the transition. In our calculations, the excited
levels of the Ne fine-structure levels are populated near-thermally at
$R=1$\,AU, but substantially sub-thermally at larger radii. The
departure from a thermal population is conveniently expressed in terms
of the factors,
\be \label{collisional_correction} 
\mathcal{C}_{ul} = 1 + \frac{n_{cr}(ul)}{n_{\rm e}}, 
\ee 
which make it possible to use Boltzmann-like formulae for the 
populations in this problem. Thus the population of the upper level 
of the Ne\,II 12.81 $\mu$m\, $J=1/2-3/2$ transition is given by, 
\be \label{12micronpop} P_{u}(1) = \frac{1}{2
\mathcal{C}_{3/2-1/2}\, e^{1122.8/T} + 1}, 
\ee and the population of the Ne\,III 15.55 $\mu$m\, is given by 
\be \label{15micronpop}
P_{u}(2) = \frac{1} {1+ 5/3\, \mathcal{C}_{1-2} \,e^{925.3/T} + 1/3\,
\mathcal{C}_{0-1} \, e^{-399/T}}.  
\ee 
Equation~\ref{15micronpop} is an approximation based on ignoring the
weak quadrupolar transitions, both collisional and radiative, that
connect the top $J=0$ and the bottom $J=2$ levels of the Ne\,III
fine-structure triplet. The factor $\mathcal{C}_{ul}$ is unity in the
high-density limit (thermal population) and proportional to the
electron density in the low-density (sub-thermal) limit where every
exciting collision produces a fine structure photon.
 
The key factors in Equation~\ref{differential_flux} for the
differential flux are the integrals,
\be
\label{integral1}
N_{u}(R;1) = \int_0^\infty dN\, P_{u}(1)\, x_1 = x_{\rm Ne}  
     \int_0^\infty dN\, P_{u}(1)\, \frac{a_1}{d},
\ee
\be
\label{integral2}
N_{u}(R;2)= \int_0^\infty dN\, P_{u}(2)\, x_2 = x_{\rm Ne} 
     \int_0^\infty dN\, P_{u}(2)\, \frac{(a_1+b)a_2}{d},
\ee
making use of Eq.~\ref{actualpops}; $N_u(R;1)$ and $N_u(R;2)$ are the
column densities of excited Ne ions that emit the dominant
fine-structure lines of Ne\,II and Ne\,III at radius $R$. Thus the
radial distribution of the flux is given by,
\be
\label{diff_flux}
dF_{ul}(n) = \frac{1}{4 \pi d^2}\, 
(A_{ul}h \nu_{ul})_n \, 
2\pi \, RN_{u}(R;n) \, dR,
\ee 
where the numerical values of $A_{ul}h\nu_{ul}$ are given in the last
column of Table 1. 

In Figure 4 we plot $N_u(R;n)$, which determines the emissivity per
unit area of the 12.81$\mu$m line of Ne$^+$ and the 15.55$\mu$m 
line of Ne$^{2+}$ through Equation~\ref{diff_flux}. Two extreme 
thermal models are shown, one for  X-ray heating dominant 
($\alpha_h = 0.01$) and the other for mechanical heating dominant 
($\alpha_h = 1.0 $). The emissivities are generally strongly peaked 
at very small radii where the upper levels are  near-thermally excited.
The effects of the central peaking is suppressed somewhat due to the factor $R$ in Equation~\ref{diff_flux} for the flux.

At larger radii, the excitation is sub-thermal, and the emissivities 
are determined by the second and third powers of the electron 
density, as suggested by Equation~\ref{chargeratios}. The electron 
density profiles decrease in magnitude roughly as the inverse square 
of the radial distance. For large $R$, the emissivity is further 
reduced by the impact of the decrease of both $\nel$ and $T$ for 
large column densities. The emissivity of the 15.55$\mu$m line is 
smaller because of the reduced abundance of Ne$^{2+}$, due in large 
part to its destruction by charge exchange with atomic hydrogen. In 
addition, the characteristic quantity $(A_{ul}h \nu_{ul})_n$ is
smaller by a factor of 1.75 for this transition compared to the
12.81$\mu$m line. The 12.81$\mu$m line of Ne$^+$ is more sensitive to
the heating model because it tends to be emitted deeper in the
atmosphere where mechanical heating can extend the warm temperature
transition region.

The unresolved line fluxes can be obtained by integrating the
emissivities in Figure 4 over radius, including the extra factor of
$R$ in Equation~\ref{diff_flux}.  The results are given in Table 2,
assuming a nominal distance for the model T-Tauri star of 
$d=140$\,pc and a Ne abundance of $10^{-4}$.
Approximately half of the flux of the 12.81$\mu$m line of Ne$^+$ 
arises from within 6\,AU, with the remaining half produced over the 
radial range $6-30$\,AU.  
\clearpage
\begin{table}
\begin{center}
\caption{Neon Ion Line Fluxes (erg s$^{-1}$ cm$^{-2}$)}
\begin{tabular}{lll}
\hline
\hline
$\alpha_h$	& Ne$^+$ 12.81 $\mu$m	& Ne$^{2+}$ 15.55 $\mu$m \\
\hline	
1.00 		& $1.25 \times 10^{-14}$ & $6.46 \times 10^{-16}$ 	\\
0.01		& $6.22 \times 10^{-15}$ & $5.27 \times 10^{-16}$     \\
\hline
\end{tabular}
\end{center}
\end{table}
\clearpage

\noindent 
The flux of the 12.81 $\mu$m line for the two extreme heating models
differs by only a factor of two, so we conclude that a typical T Tauri
star generates a flux of $\sim 10^{-14} \erg\, \psqcm \ps$; the flux
of the 15.55 $\mu$m line is about 20 times smaller, $\sim 5\times
10^{-16} \erg \, \psqcm \ps$. The predicted value of the flux of the
12.81 $\mu$m line suggests that it may be observable with only
moderate spectral resolution. The dust continuum emission for the
D'Alessio generic T-Tauri star (at 140\,pc) near $13 \micron$ is $1.21
\times 10^{-11}$ erg s$^{-1}$ cm$^{-2}$ $\mu$m$^{-1}$. If we consider
the {\it Spitzer} Infrared Spectrometer (IRS), with a spectral
resolution of $600$, the continuum flux within a spectral resolution
element is $2.58 \times 10^{-13}$ erg s$^{-1}$ cm$^{-2}$, about 25
times larger than the theoretically predicted line flux. Thus a high
signal-to-noise observation with the IRS should lead to a detection of
the 12.81 $\mu$m line. The Ne$^{2+}$ 15.55 $\mu$m line would be much
more difficult to detect. These lines would become easier to detect in
sources with weaker continua or with observations made at higher
spectral resolution. Because H atom collisions have been ignored, the fluxes in Table 1 should be considered as lower limits.


This discussion of the observability of the Ne fine-structure lines
has been based on a neon abundance of $10^{-4}$, or equivalently on a
solar abundance ratio of neon to oxygen, Ne/O = 1/6. Several lines of
evidence have suggested that the Ne abundance is closer to $x_{\rm Ne} 
= 2.5 \times 10^{-4}$. {\it Chandra} X-ray spectroscopy of nearby stars with a wide range of the ratio $\LX /\Lbol$ give a nearly constant Ne/O =
0.4 (Drake \& Testa 2005). For the few T Tauri stars where the same measurements have been made, two are in harmony with this result (BP Tau 
and TWA 5 with Ne/O $\approx 0.5$ ) and one has an even larger value 
(TW Hya with Ne/O $= 0.87 \pm 0.13$), as discussed by Drake, Testa \& Hartmann (2005).  Measurements of Orion B stars also give a larger Ne abundance than $10^{-4}$, $x_{\rm Ne} = 1.29 \times 10^{-4}$ and a 
larger neon to oxygen ratio, Ne/O = 0.25 (Cunha, Hubeny \& Lanz 2006).  
The solar Ne abundance plays a key role in the continuing controversy between the results of helio-seismology and solar models (Asplund \etal 2004, Antia \& Basu 2005, Bahcall, Basu \& Serenelli 2005, Schmelz \etal 2005, Delahaye \& Pinsonneault 2006). 
Despite these uncertainties, our results suggest that disk Ne 
abundances anywhere in the above range of values are sufficient to 
produce detectable Ne$^+$ emission. 

	The spatial distribution of the emissivity in Figures 4 has
consequences for the rotational broadening of the Ne fine structure
lines emitted by protoplanetary disks. Assuming Keplerian rotation,
the column density $N_u(R;n)$ can be converted into a distribution
function $P(v)$ for the rotational speed 
($P(v)dv \propto N_u(R) R\,dR$) 
such that $P(v)dv$ is the fraction of emission within the speed interval $dv$,
\be
\label{velocity_distribution}
P(v) \propto v^{-5} N_u(R(v);n) 
\ee
where $R= G \Mst v^{-2}$. When combined suitably with the distribution
of turbulent velocities, $P(v)$ helps determine the shape of the
lines. Figure 5 shows the distribution function $P(v)$ for the Ne\,II
12.81\,$\micron$ line, where the abscissa is $w=v/v(1\,{\rm AU})>
0$. We see that $P(v)$ has two components. One is a narrow
distribution centered at $0.25\, v(1\,{\rm AU})$, which corresponds to
the cut-off of the spatial variation of $N_u(R;n)$ in Figure 4 beyond
16\,AU. Then there is a long tail extending to the maximum rotational
speed, which arises from the peaking of the spatial distributions in
Figure 4 at small $R$. Ignoring turbulent broadening, the rotational
lineshape function for inclined disks will have a double peak, associated with the cut-off of the emissivity at large radii, plus extended line wings associated with the concentration of the emission at small radii. In principle, this analysis can be used to obtain information about the
spatial distribution of the emission from measurements of the
lineshape. Assuming that the turbulent velocities are of order $1 \,
\kmps $, as suggested by the measured CO lineshapes of T Tauri stars
(Najita \etal 2006), moderately-high spectral resolution observations
with $R \sim 3 \times 10^4$ could confirm the expectations for the Ne
lineshapes based on Figure 5.

\section{Discussion} 

In the last section, we concluded that the $\nep \, 12.81 \micron$
line should be detectable from a typical T Tauri star with a moderate
resolution spectrometer like the {\it Spitzer} IRS. Indeed. we have
reports from colleagues of the detection with this instrument of
unusually strong $12.81 \micron$ emission from a few T Tauri stars
(Ilaria Pascucci 2006, Dan Watson 2006 private communications). These
are probably the brightest Ne fine-structure emission line sources;
the emission from most others is likely to be fainter. This line is
also observable from the ground, and attempts to observe it with a
large telescope at high spectral resolution would be of great
interest. Not only would the line to continuum ratio be increased, but
it might become possible to obtain information about the line shape,
as discussed at the end of \S 3, that would help locate the source of
the emission in the disk.

We have argued that the detection of the Ne ion fine-structure lines
would provide clear support for the role of stellar X-rays in
determining the physical properties of the atmosphere of YSO disks,
the region most accessible to observational study. The Ne lines
complement the detection of the ro-vibrational transitions of CO and
the UV fluorescence of $\hm$ (reviewed by Najita \etal 2006). The
latter transitions are associated with material at disk radii within a
few AU and with vertical column densities $\Nh \gtrsim 10^{21}
\psqcm$. Our calculations show that the emission of the Ne fine
structure lines arises from a larger range of radial distances, out to
$10-15$\,AU, and that the lines are formed mainly at vertical column
densities in the range $\Nh \sim 10^{19}- 10^{20} \psqcm$. Thus the Ne
fine-structure transitions are complementary to the lines of CO and
$\hm$ in probing both the warm temperature region above the cool
molecular layer and a region that extends to relatively large radii
that, in our own solar system, would range from the location of the
terrestrial planets to beyond the giant planets.  By contrast, both
the UV transitions of H$_2$ and the CO ro-vibrational transitions are
generally restricted to the inner few AU of disks (Najita \etal 2003;
Najita \etal 2006), and the IR transitions of H$_2$ are likely to probe a
narrow range of radii (Pascucci \etal 2006).


The measurement of Ne lines in the disks of low-mass YSOs is also of
interest because of the special chemistry of neon. From the disk
modeling perspective, Ne has the advantage that the complexities of
molecule formation and destruction do not enter. It is unlikely that
significant amounts of Ne will freeze out on dust grains, although
some of the neon trapped in meteorites has been interpreted in terms
of adsorption (e.g., Swindle 1988). Thus the detection of the Ne
fine-structure lines, in conjunction with model calculations, might
eventually provide information on the abundance of Ne in YSO disks,
which serve as observable analogs of the primitive solar
nebula. Because the Ne lines arise in a relatively thin layer at high
altitudes, by themselves they provide only weak constraints on the
total gas mass of protoplanetary disks. However, when taken together
with other diagnostics, such as CO and $\hm$, detailed modeling of
several species might provide important information on disk gas mass.

The calculations reported in this paper are limited to the
fine-structure lines of just one element and one type of disk, the
typical T Tauri disk as formulated by D'Alessio \etal (1999). In
addition to extending the range of application of models such as ours
(see also the work of Gorti \& Hollenbach 2004), attention needs to be
given to the underlying physics and chemistry of disk modeling, some
of which is quite uncertain and affects the accuracy of model predictions.

The emission of Ne fine-structure lines from relatively cool YSO disks
arises from the ejection by hard X-rays of the K-shell electrons of
moderately heavy ions, followed by the Auger effect. This process is
enhanced as the mass of the target ion is increased, and it surely
occurs for other abundant elements such as Si, S, and Fe. Our analysis
of Ne fine-structure lines shows how important the recombination
cascade from higher to lower ion stages is in determining the
abundance of the first few (and most abundant) ions. Sulfur serves as
a good example. Gorti and Hollenbach (2004) and Hollenbach \etal
(2005) have suggested that the S\,I 25.2 and 56.3 $\micron$
fine-structure lines, among others, are potential diagnostics of
intermediate-age YSO disks. Hard X-rays will produce ions as high as
S$^{6+}$, and the highest ions are expected to be destroyed by a chain
of fast charge transfers with atomic hydrogen, as has been confirmed
by a high-level theoretical calculation by Stancil \etal (2001) of the
reaction S$^{4+}$ + H $\rightarrow$ S$^{3+}$ + H$^{+}$, which is
fast down to low energies. However, the situation for S$^{3+}$ and
S$^{2+}$ remains problematic. Early calculations of S$^{2+}$ + H
$\rightarrow$ S$^{+}$ + H$^{+}$ (Butler \& Dalgarno 1980, Christensen
\& Watson 1981) suggest that the rate coefficient is small at low
temperatures. Recent calculations of S$^{3+}$ charge exchange
(Bacchus-Montabonel 1998; Labuda \etal 2004) are restricted to keV
energies.  The recent study of the reaction, H$^{+}$ + S $\rightarrow$
S$^{+}$ + H, by Zhao \etal (2005) provides important information for
the ionization balance between S and S$^+$, but theoretical and
experimental studies of S$^{2+}$ and S$^{3+}$ charge transfer with H
at low energies ($\lesssim 1$\,eV) are needed before the emission of
the S\,I and S\,III fine-structure lines can be calculated reliably
for X-ray irradiated disks, e.g., along the lines of the analysis for
neon in this paper. A similar conclusion applies to other heavy
elements for which low-energy charge-transfer cross sections with H
are lacking for the first few ions.

It is important to recall the discussion of low-energy Ne ion charge
exchange in \S 2. In particular, we have relied on 25 year old
theoretical estimates in adopting a rate coefficient $k_2 = 10^{-14}
\ccps$ for the charge transfer from Ne$^{2+}$ to H.  Because the
detectability of the Ne\,III 15.55 $\micron$ line depends on this
coefficient, it would be important to have more theoretical work on
this reaction. We also have no information on the rate coefficients
for the excitation of the Ne ion fine-structure levels by atomic
hydrogen. Although few in number, existing calculations (e.g., Launay
\& Roueff 1977 \& Barinovs et al. 2005 for H + C$^+$) lead to rates
$\sim 10^{-9}$cm$^3$s$^{-1}$. If this order of magnitude is
appropriate for the Ne ions, then H atom collisions would be important
in that part of the disk atmosphere where $\xel < 0.01$ and $T$ is
still large enough to excite levels with excitation temperatures of
order 1000\,K. Our flux estimates may also be affected by the lack of
understanding of other basic microscopic processes, not to mention the
many astrophysical uncertainties such as disk structure, dynamics and
thermodynamics. We have considered the possibility that the Ne ions
are destroyed by molecules and PAHs.  It is known from room
temperature experiments (Anicich 1993) that $\nep$ does not react
strongly if at all with $\hm$ and CO, molecules that are predicted to
occur at reduced levels in the warm upper atmosphere. Depending on
their charge and abundance, PAHs might conceivably pay a role in
destroying Ne ions and thus reduce the strength of the fine-structure
lines. Negatively charged PAHs can neutralize Ne ions with large
reaction rate coefficients (e.g., Tielens 2005).
Both capture and charge transfer are possible reactions with neutral
PAHs. Using rate coefficient estimates in the literature (Bohme 1992,
Tielens 2005) and a maximal PAH abundance of $10^{-7}$, we find that
these neutral reactions are less important than destruction by
radiative recombination. Of course, there is also the strong
possibility that the PAH abundance in YSO disk atmospheres is reduced
by the strong stellar X-ray flux or by the ions that the X-rays
generate. It would appear that the largest uncertainty in the
underlying microphysics of our analysis of the neon ion abundances and
their fine-structure line emission is in the magnitude of the weak but
not insignificant charge transfer to atomic hydrogen.

One astrophysical uncertainty that affects our estimates of the line
fluxes is the GNI04 choice of the X-ray luminosity and spectral
temperature, $\LX = 2\times 10^{30}$\,erg\,$\ps$ and $k\TX = 1$\,keV.
The former is characteristic of solar-like YSOs observed by {\it
Chandra} in the Orion Nebula Cluster (Garmire \etal. 2000, Wolk \etal
2005). This may well be an overestimate for the X-ray luminosity of
lower-mass and older T Tauri stars.  Based again on {\it Chandra}
observations of the Orion Nebula Cluster (Preibisch \etal 2005), the
X-ray luminosity is measured to be smaller for lower-mass and older
YSOs, and it is suppressed somewhat for accreting as opposed to
non-accreting systems.  Rather than use just a single temperature, it
would be more appropriate to represent the X-ray spectrum with a
two-temperature model of the type used to fit X-ray observations of
well-observed stars. More generally, in future modeling of specific
stars, the observed X-ray spectra should be used.

We can use the approximations developed in \S 2, especially
Equations~\ref{simple_ionization} and ~\ref{chargeratios}, to identify
some of the other astrophysical uncertainties in our flux
estimates. The Ne levels are sub-thermally excited over most of the
disk volume responsible for the emission of the fine-structure
lines. Thus the local emissivity of the Ne\,II 12.81\,$\micron$ line
is determined by $\nel ^2$, because the $\nep$ abundance usually
tracks the electron fraction
(Eq.~\ref{chargeratios}). Equation~\ref{simple_ionization} now tells
us that the emissivity is determined roughly by \be
\label{essential_factors}
j(\nep) \propto x_{\rm Ne} \, \nh \, \zeta({\rm Ne}) \,
\frac{k_{lu}(T)}{\alpha_1(T)}, \ee i.e., by the product of the Ne
abundance, the local density, the ionization rate (including
attenuation), and a function of temperature that is essentially the
Boltzmann factor, $\exp{(-T_{ul}/T)}$. This last factor is one of
several responsible for cutting off the emissivity at large column
densities (and large radii). Equation~\ref{essential_factors} suggests
that the Ne\,II 12.81\,$\micron$ line strength will be larger for
younger and more active T Tauri stars, which have larger densities and
ionization rates. However, this does not mean that the line will be
easier to detect from such stars, because their dust continuum
emission will also be larger.

Flaring of the disk also plays a role in that the more flaring the
smaller the attenuation of the X-rays. The GNI04 is flawed in this
respect because it uses the D'Alessio \etal (1999) density
distribution, whose scale height is determined by the dust
temperature. In a thermal-chemical model that treats hydrostatic
equilibrium self-consistently, we would expect an even more diffuse
atmosphere than the one used in this paper. Furthermore, Table 2 shows
that, without mechanical heating, the flux Ne\,II 12.81\,$\micron$
line is reduced by a factor of two. In other words, the emissivity is
affected by the poorly understood processes that heat the disk
atmosphere. Clearly, there are many possibilities for variation of the
Ne fine-structure line fluxes, and some if not many T Tauri stars may
generate fluxes smaller than those given in Table 2.

In conclusion, we have modeled the physical properties of the disk
atmosphere of a generic T Tauri star exposed to a strong stellar X-ray
flux, and we have calculated the abundance of $\nep$ and Ne$^{2+}$ and
the strength of their fine-structure lines. The estimated fluxes
indicate that the Ne\,II 12.81\,$\micron$ line should be detectable
with existing instrumentation. The observation of this line from warm
disk atmospheres would provide strong evidence for the significant
role of stellar X-rays on the physical and chemical properties of YSO
disks and on the conditions for photoevaporation.

\acknowledgements
The authors would like to thank Alex Dalgarno for advice on neon ion
charge transfer reactions with atomic hydrogen, Phil Stancil for
information about sulfur ion charge exchange, Jack Lissauer for
discussions on the potential role of neon in determining the gas mass
of protoplanetary disks (a key quantity for the formation of the
jovian planets) and David Hollenbach for discussing the possible role of EUV radiation in generating Ne ions and especially for his many helpful comments on this manuscript.  This research has been supported by
grants from NSF and the NASA Origins Program.

{}

\clearpage

\figcaption{Temperature profiles for a protoplanetary disk atmosphere
based on GNI04. The radial distance is 5\,AU, and the mass accretion
rate is $10^{-8} \Msun$ yr$^{-1}$. The abscissa is vertical column
density in cm$^{-2}$, and the ordinate is temperature in degrees
K. The lower dashed line is the dust temperature for the D'Alessio et
al.\ (1999) generic T Tauri star model. The upper solid curves are gas
temperature labeled by the phenomenological mechanical surface heating
parameter defined by Equation~\ref{accheat}: $\alpha_h = 1$ (mechanical 
heating dominant) and $\alpha_h = 0.01$ (X-ray heating dominant).  
The major chemical transitions occur between
$10^{21}$ and $10^{22}\, {\rm cm}^{-2}$: significant levels of $\hm$
and full formation of CO occur near $\Nh \sim 1-2 \times 10^{21} {\rm
cm}^{-2}$, and complete formation of $\hm$ is achieved at $\Nh = 8.5
\times 10^{21}\, {\rm cm}^{-2}$.}

\figcaption{Gas temperature profiles for the GNI04 disk atmosphere
model for the case $\alpha_h = 0.01$. The abscissa is vertical column
density in cm$^{-2}$, and the ordinate is temperature in degrees
K. The general structure consists of a thin hot layer on top of a
thick cool layer (near the midplane), with a warm transition layer in
between.}

\figcaption{Electron fractions for the GNI04 X-ray irradiated disk
atmosphere model for the case $\alpha_h = 0.01$. The abscissa is
vertical column density in cm$^{-2}$, and the ordinate is $\xel$, the
electron abundance relative to total hydrogen. Deeper down than a
vertical column $\Nh = 3 \times 10^{20} \psqcm$, the electron fraction
is $\lesssim 10^{-4}$.}

\figcaption{The Ne fine-structure line emissivity per unit area,
expressed in terms of column densities of the upper level from
Equation~\ref{diff_flux}, calculated with the GNI04 disk atmosphere
model for a neon abundance of $x_{\rm Ne} = 10^{-4}$. There is a pair
of curves for each fine-structure transition, Ne\,II $12.81 \micron$
and Ne\,III $15.55 \micron$. The upper curve is for $\alpha_h = 1$
(mechanical heating dominant), and the lower curve is for $\alpha =
0.01$ (X-ray heating dominant).  The emissivity is peaked at small
$R$, and then drops off rapidly beyond $R \sim 16$\,AU. }

\figcaption{The distribution of rotational speeds of the Ne\,II 12.81
$\micron$ line emission. The abscissa is speed, normalized to the
value at 1\,AU (the smallest radius used in the present emissivity
calculation), and the ordinate is the unnormalized probability. The
peak is located near $0.25\,v(1{\rm AU})$, corresponding to the
large-$R$ cut-off in Figure 4. The long tail at large velocities is
associated with the strong emission at small radii. This distribution,
convolved with the distribution of turbulent velocities, determines
the line shape, which will be double-peaked and exhibit substantial
line wings for inclined disks.}

\clearpage

\begin{figure}
\plotone{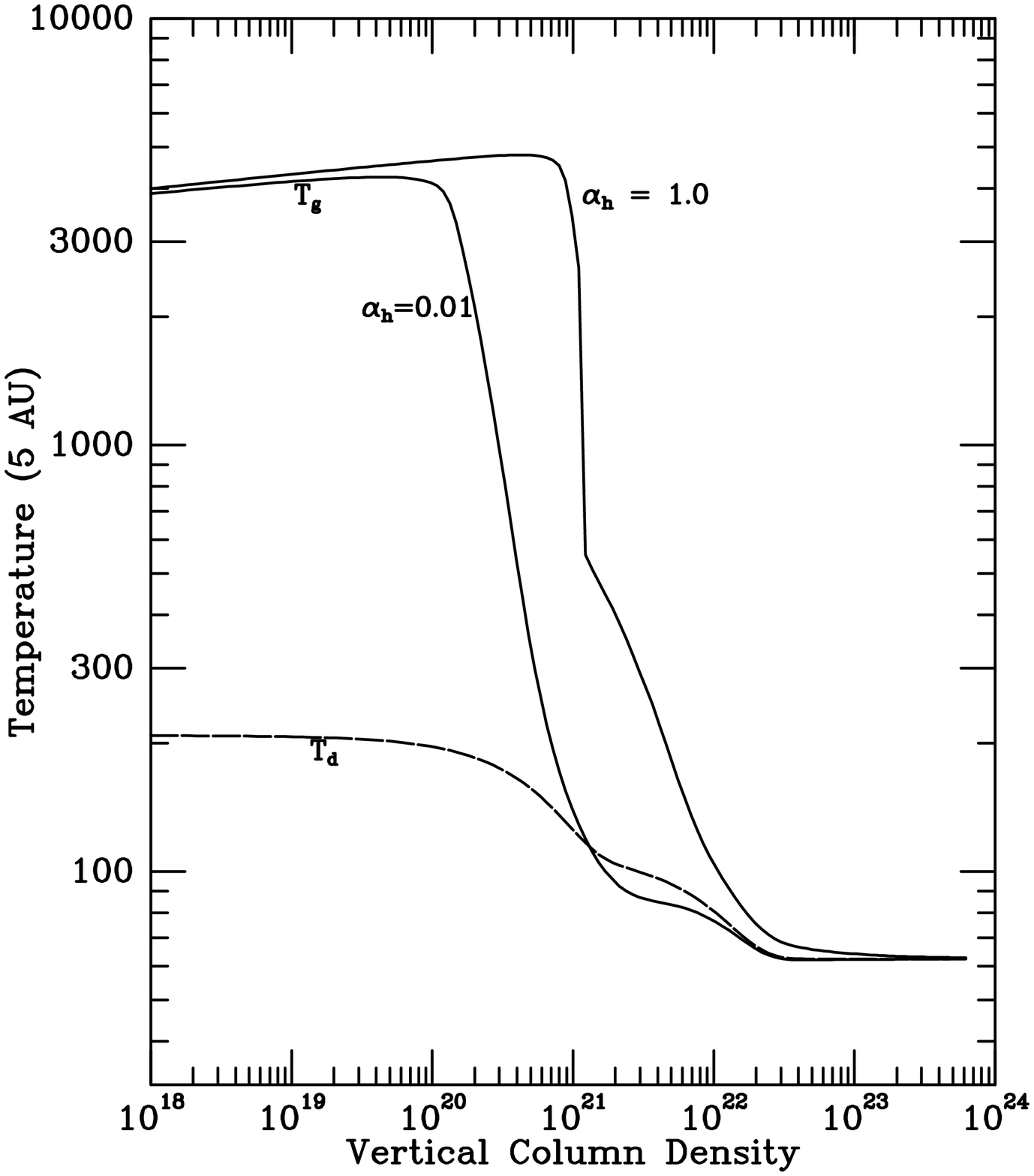}
\centerline{f1.eps}
\end{figure}
\clearpage

\begin{figure}
\plotone{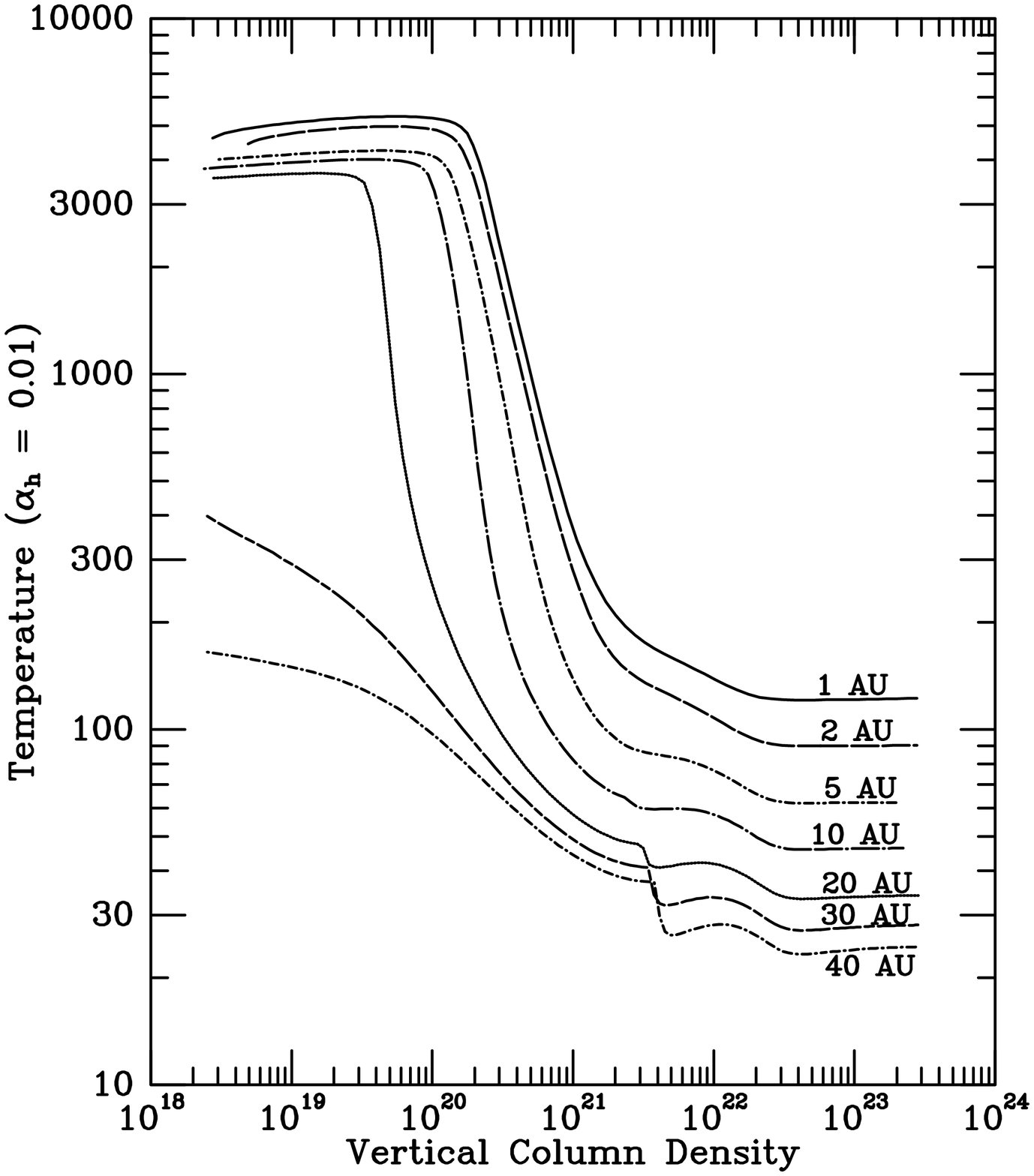}
\centerline{f2.eps}
\end{figure}
\clearpage

\begin{figure}
\plotone{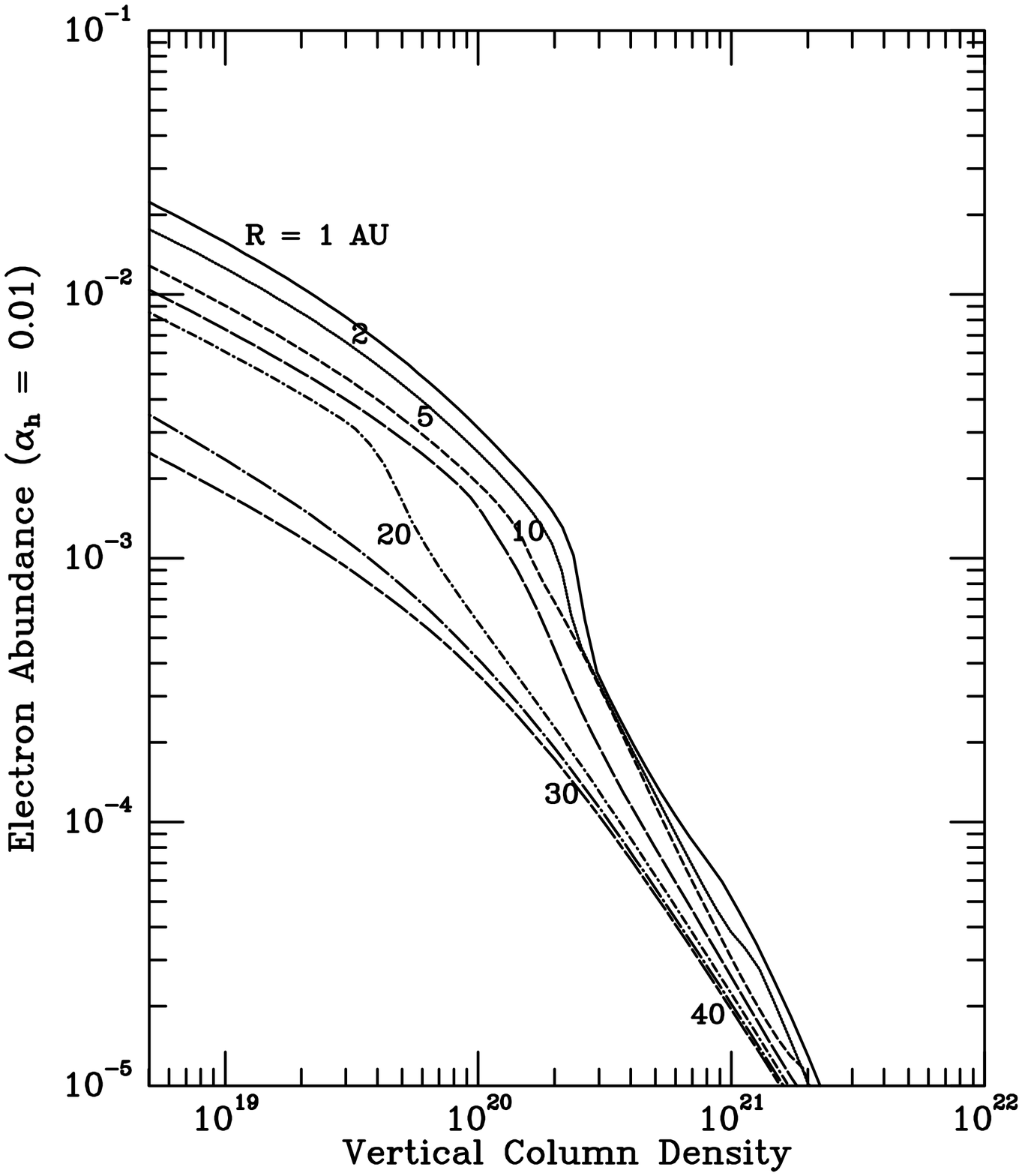}
\centerline{f3.eps}
\end{figure}
\clearpage

\begin{figure}
\plotone{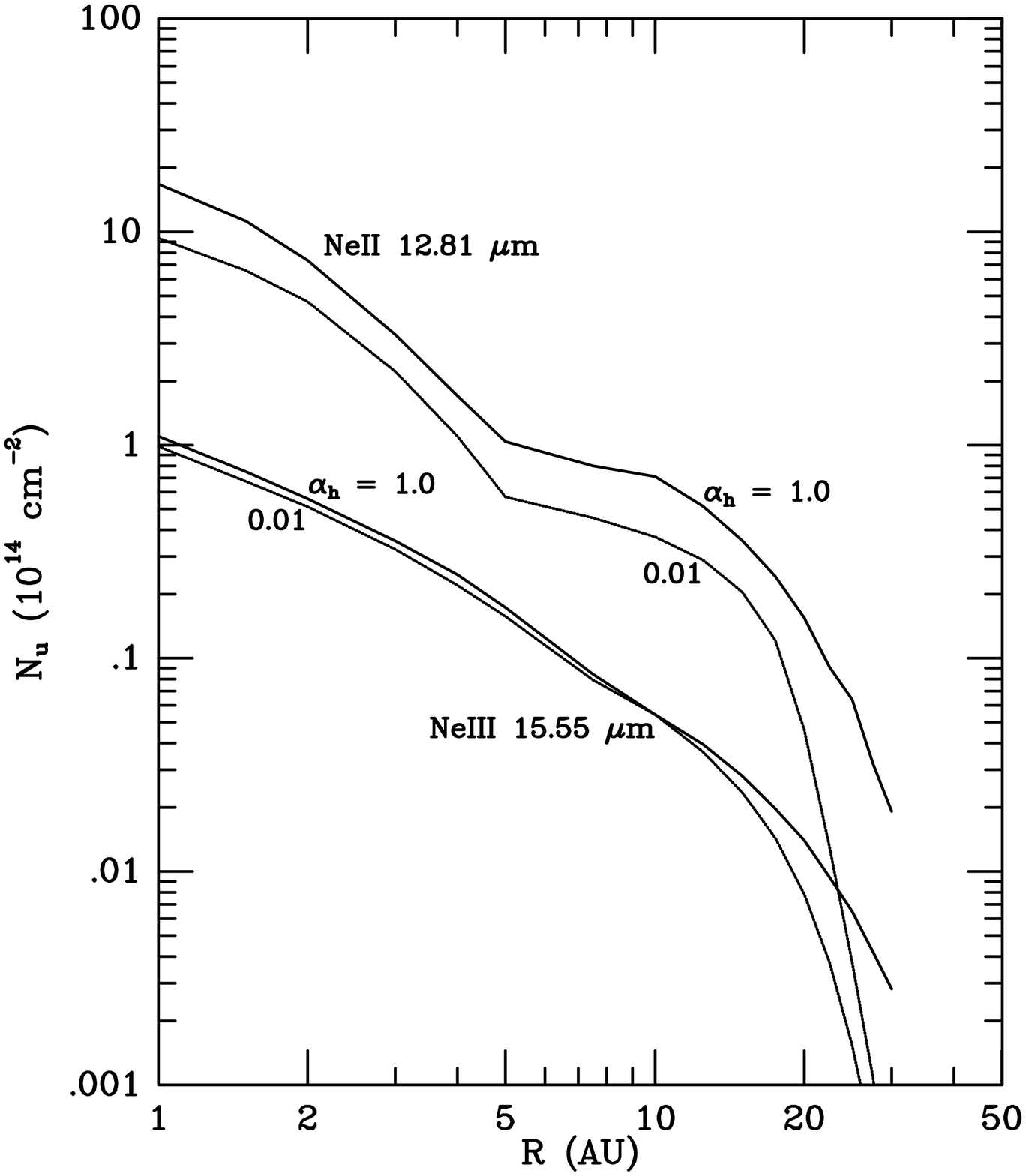}
\centerline{f4.eps}
\end{figure}
\clearpage

\begin{figure}
\plotone{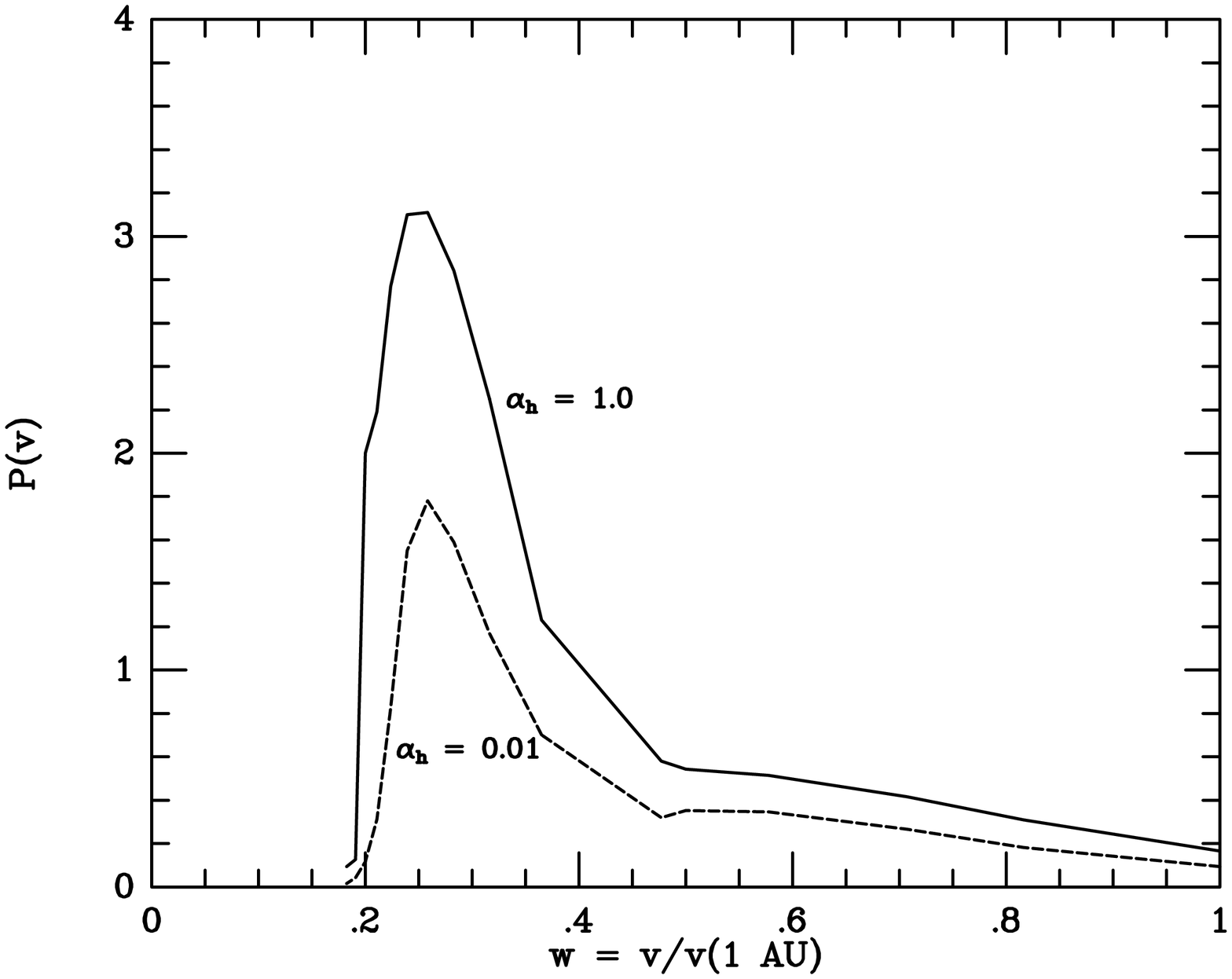}
\centerline{f5.eps}
\end{figure}

\end{document}